\documentclass[aps,pre,numerical,superscriptaddress,showpacs,preprint]{revtex4-1} 

\usepackage{bm}
\usepackage{amsmath}
\usepackage{epsfig}
\usepackage{rotating}
\usepackage{graphicx}
\usepackage{hyperref}

\def\pr{\prime}
\def\be{\begin{equation}}
\def\lan{\left\langle}
\def\ran{\right\rangle}
\def\ee{\end{equation}}
\def\barr{\begin{array}}
\def\earr{\end{array}}

\def\l{\left}
\def\r{\right}
\def\dis{\displaystyle}
\def\ed{\end{document}}

\def\can{{\cal N}}

\begin{document}

\title{Quenched many-body quantum dynamics with $k$-body interactions using $q$-Hermite polynomials}

\author{Manan Vyas}
\thanks{corresponding author, manan@icf.unam.mx}
\affiliation{Instituto de Ciencias F{\'i}sicas, Universidad Nacional 
Aut{\'o}noma de M\'{e}xico, 62210 Cuernavaca, M\'{e}xico}
\author{V. K. B. Kota}
\affiliation{Physical Research Laboratory, Ahmedabad 380 009, India
}

\begin{abstract}
In a $m$ particle quantum system, the rank of interactions and the nature of particles (fermions or bosons) can strongly affect the dynamics of the system. To explore this, we study non-equilibrium dynamics with the particles in a one-body mean-field and quenched by an interaction of body-rank $k=2$, $3$, $\ldots$, $m$. Using Fermionic Embedded Gaussian Orthogonal Ensembles (FEGOE) and Bosonic Embedded Gaussian Orthogonal Ensembles (BEGOE) of one plus $k$-body interactions (also the Unitary variants FEGUE and BEGUE), it is seen that the short time decay of the survival probability of many-particle systems is given by the Fourier transform of the generating function $v(E|q)$ of $q$-Hermite polynomials. Deriving formulas for $q$ for both fermion and boson systems as a function of $m$, $k$ and number of single particle states $N$, we have verified that the Fourier transform of $v(E|q)$ agrees very well with numerical ensemble calculations for both fermion and boson systems. These results bridge the gap between the known results for $k=2$ and $k=m$.
\end{abstract}


\maketitle

\section{Introduction}

In the last decade, there is special focus on statistical properties of
isolated finite many-particle quantum systems such as atomic nuclei, atoms,
mesoscopic systems (quantum dots, small metallic grains), interacting spin
systems modeling quantum computing core, ultra-cold atoms and quantum black
holes with SYK model and so on \cite{kota, zel-lea, Rigol, Ko-17, bh-1}. A route to
investigate statistical properties of isolated finite many-particle quantum
systems is to employ the classical Gaussian orthogonal (GOE) or unitary (GUE) or
symplectic (GSE) random matrix ensembles with various deformations
\cite{RMT-book,kota}. However, in most of the isolated finite many-particle
quantum systems, their constituents predominantly interact via few-particle
interactions and one refinement of the classical ensembles which retains the
basic stochastic approach but allows for this feature consists in the use of
embedded random matrix ensembles \cite{Br-81, BW, Man-th, kota, Ko-17, Ma-18}.

Representing an isolated finite interacting quantum system, say with $m$
particles (fermions or bosons) in $N$ single particle (sp) states, by random
matrix models generated by random $k$-body interactions and propagating the
information in the interaction to many particle spaces, we have random
interaction matrix models for $m$-particle systems. In the simplest version, the
$k$-particle Hamiltonian ($H$) of a spinless fermion (or boson) system is
represented by GOE/GUE/GSE (all three classical ensembles combined are referred
as GE, Gaussian ensembles) and then the $m$ particle $H$ matrix is generated
using the $m$-particle Hilbert space geometry. As a GOE/GUE/GSE
random matrix ensemble in $k$-particle spaces is embedded  in the $m$-particle
$H$ matrix, these ensembles are generically called $k$-body embedded ensembles
[EE($k$)] \cite{Fr-71, Bo-71}. Then, with GOE embedding, we have embedded Gaussian orthogonal ensemble of $k$-body interactions [EGOE($k$)] and similarly with GUE embedding EGUE($k$). In order to distinguish fermion and boson systems, following \cite{Ma-18}, we will use the notation FEGOE($k$) and FEGUE($k$) for fermionic systems (here, F denotes fermions) and BEGOE($k$) and BEGUE($k$) for bosonic systems (here, B denotes bosons). For common reference, we will use EE$(k)$ that includes all four mentioned cases. 

Following the seminal paper of Mon and French \cite{MF} and many numerical
calculations \cite{Br-81,kota,Manan,Ko-17}, it is well known that the FEGOE($k$) and
FEGUE($k$) spectral density for a system of $m$ spinless fermions (also for
BEGOE($k$) and BEGUE($k$) for a system of $m$ spinless bosons) in $N$ sp states changes from Gaussian to the semi-circle of classical RMT as the body rank $k$ of the interaction changes from $k=1$ to $k=m$. This is also proved later by evaluating lower order moments of the spectral density by many other groups using different methods \cite{BW, Asaga, kota, Mu-16}. However, the most recent study of the spectral density of the so called SYK model \cite{Verb-1, Verb-2, Verb-3} and quantum spin glasses \cite{sping-1} employs $q$-Hermite polynomials. We show that generating function for $q$-Hermite polynomials describes the semi-circle to Gaussian transition in spectral densities and local density of states (LDOS) [also known as strength functions] of FEGOE$(k)$ and BEGOE$(k)$ (also the Unitary variants FEGUE$(k)$ and BEGUE$(k)$) as a function of rank of interactions $k$. LDOS gives the spread of the basis states over the eigenstates. It is important to mention that the spectral densities also exhibit transition from semi-circle to Gaussian form as one increases number of particles $m$ with a given $k$; see \cite{Br-81, kota} for more details. However, we do not deal with this situation in the present paper. 

Thermalization of isolated finite interacting quantum systems is a topic of great current interest \cite{Rigol}. In this context, study of time evolution of a many-body quantum system quenched far from equilibrium has attracted considerable attention and the survival probability decay is a fundamental quantity of interest in all investigations \cite{Lea, Haldar, Manan}. In \cite{Manan}, a first attempt is made to obtain analytical and numerical results for survival probability decay for fermions in a mean-field quenched by a random $k$-body interaction with $k$ changing from $1$ to $m$. The Hamiltonian with a mean-field one-body part and a $k$-body interaction represented by FEGOE($k$) with some strength $\lambda$ form FEGOE$(1+k)$ and for sufficiently large value of $\lambda$, the FEGOE$(1+k)$ properties go over to those of FEGOE$(k)$ \cite{kota, Manan}. For chaotic systems, analytical results for survival probability decay using FEGOE($k$) are derived  \cite{Manan} in the two extreme limits of Gaussian LDOS (here $k << m$ applies) and semi-circle form for LDOS (then full random matrix or GOE and equivalently FEGOE($k$) with $k=m$ applies). We demonstrate that the numerical Fourier transform of the generating function of $q$-Hermite polynomials describes the short time decay of survival probability in FEGOE$(1+k)$ and BEGOE$(1+k)$. We derive the formula for $q$ for FEGOE($k$), FEGUE($k$) and BEGUE($k$) [also for BEGOE$(k)$] that governs the behavior of spectral density, LDOS and survival probability.  Now, we will give a preview.

In Section \ref{sec1}, some basic properties of $q$-Hermite polynomials and their generating function are described. We also define embedded ensembles for finite interacting quantum systems. Using these, formulas for the parameter $q$ are presented for FEGOE($k$), FEGUE($k$) and BEGUE($k$) in Section \ref{sec2}. These results are tested for spectral densities using numerical examples for FEGOE$(k)$, FEGUE$(k)$, BEGOE$(k)$ and BEGUE$(k)$ in Section \ref{sec3}. Section \ref{sec4} gives results for LDOS and survival probability decay in FEGOE(1+$k$) and BEGOE(1+$k$) which are compared to the generating function of $q$-Hermite polynomials and its numerical Fourier transform respectively. Finally, Section \ref{sec5} gives conclusions. 

\section{Preliminaries}
\label{sec1}

\subsection{$q$-Hermite polynomials}

The $q$-Hermite polynomials were introduced by L. J. Rogers who used them to prove the Rogers-Ramanujan identities \cite{qherm-1}. It is well known in mathematics literature that $q$-Hermite polynomials are orthogonal with respect to a function that takes Gaussian form for $q=1$ and semi-circle form for $q=0$ \cite{qherm-1, qherm-2}. We will
restrict our discussion to $q$ real. In this section, we collect some basic properties of $q$-Hermite polynomials and then use them to describe the spectral density and survival probability of EE($k$).

Let us begin with the $q$ number $[n]_q$ defined by
\be
\l[n\r]_q = \dis\frac{1-q^n}{1-q} = 1+q + q^2 + \ldots+q^{n-1}\;.
\label{eq.qh1}
\ee
Note that $[n]_{q \rightarrow1}=n$. Similarly $[n]_q!=\Pi^{n}_{j=1} [j]_q$ with
$[0]_q!=1$. Now, $q$-Hermite polynomials $H_n(x|q)$  are defined by the recursion
relation \cite{qherm-1, qherm-2},
\be
x\,H_n(x|q) = H_{n+1}(x|q) + \l[n\r]_q\,H_{n-1}(x|q)
\label{eq.qh2}
\ee
with $H_0(x|q)=1$ and $H_{-1}(x|q)=0$. Note that for $q=1$, the $q$-Hermite
polynomials reduce to normal Hermite polynomials (related to Gaussian) and for
$q=0$ they will reduce to Chebyshev polynomials (related to semi-circle). More
importantly, $q$-Hermite polynomials are orthogonal within the limits $\pm
2/\sqrt{1-q}$, with the weight function $v(x|q)$ (see Eq. (2.14) of 
\cite{qherm-1}),
\be
\dis\int^{2/\sqrt{1-q}}_{-2/\sqrt{1-q}} H_n(x|q)\,H_m(x|q)\,v(x|q)\,dx =
\l[n\r]_q!\,\delta_{mn}\;.
\label{eq.qh3}
\ee
Explicit form of $v(x|q)$ is given by Eq. (2.15) of Ref. \cite{qherm-1}. After 
some simplifications of this equation, it is easy to see that
\be
\barr{rcl}
v(x|q) & = & \can_q\,\dis\sqrt{1-\dis\frac{x^2}{x_0^2}}\;
\dis\prod^{\infty}_{\kappa = 1} \l[
1-\dis\frac{4(x^2/x_0^2)}{2+q^\kappa + q^{-\kappa}}\r]\;;\;\;
\\
x_0^2 & = & \dis\frac{4}{1-q}\;.
\earr
\label{eq.qh4}
\ee
Here, $x$ is standardized variable (has zero mean and variance unity) with
$-2/\sqrt{1-q} \leq x \leq 2/\sqrt{1-q}$ and $\can_q$ is a normalization
constant such that $\int^{2/\sqrt{1-q}}_{-2/\sqrt{1-q}} v(x|q)\,dx = 1$. It is
seen that in the limit $q \rightarrow 1$, $v(x|q)$ will take Gaussian form and
in the $q=0$ limit, $v(x|q)$ will be semi-circle. Thus, $v(x|q)$ interpolates 
Gaussian and semi-circle forms. It is also shown in \cite{qherm-1} (see 
proposition 4.1 and its proof) that the even order reduced moments of $v(x|q)$ are ,
\be
\barr{rcl}
\mu_{2n}(q) & = & \dis\int^{2/\sqrt{1-q}}_{-2/\sqrt{1-q}} x^{2n}\,v(x|q)\,dx \\
&  = & (1-q)^{-n}\,\dis\sum_{r=-n}^{r=n} {2n \choose n+r} (-1)^r\,
q^{r(r-1)/2}\;.
\earr \label{eq.qh5}
\ee
Note that all the odd order moments vanish. Simplifying Eq. (\ref{eq.qh5}) for
$n=2$, 3 and $4$ will give the following important formulas (it is easily seen
that $\mu_2=1$ as we are using standardized variable),
\be
\barr{l}
\mu_4(q) = 2+q\;,\\
\mu_6(q) = 5+6q+3q^2+q^3\;,\\
\mu_8(q) = 14+28q+28q^2+20q^3+10q^4+4q^5+q^6\;.
\earr \label{eq.qh6}
\ee
Using these formulas and the moments to order 8 for FEGUE($k$), FEGOE($k$), BEGUE($k$), and BEGOE($k$), it is seen that the spectral densities and LDOS will be close to $v(x|q)$ that generates $q$-Hermite polynomials. With this, the formulas for $q$ are identified in the next section. 

\subsection{Embedded ensembles}

Constituents of finite many-body quantum systems such as nuclei, atoms, molecules, small metallic grains, quantum dots, arrays of ultracold atoms, and so on, interact via few-body (mainly two-body) interactions. As is well-known, the classical random matrix ensembles (GOE/GUE) incorporate many-body interactions. Embedded ensembles take into account the few-body nature of interactions and hence, they are more appropriate for analyzing various statistical properties of finite quantum systems.

Given a system of $m$ particles (fermions or bosons) distributed in $N$ degenerate levels interacting via $k$-body $(1 \leq k \leq m)$ interactions, embedded ensembles are generated by representing the few particle ($k$) Hamiltonian by a classical GOE/GUE and then the many-particle Hamiltonian ($m>k$) is generated by the Hilbert space geometry. In other words, $k$-particle Hamiltonian is embedded in the $m$-particle Hamiltonian and the non-zero $m$-particle Hamiltonian matrix elements are appropriate linear combinations of the $k$-particle matrix elements. Due to the $k$-body selection rules, many matrix elements of the $m$-particle Hamiltonian will be zero unlike in a GOE.

The random $k$-body Hamiltonian in second quantized form for a FEGOE/BEGOE ($\beta=1$) and FEGUE/BEGUE ($\beta=2$) is \cite{Ma-15, Ma-18},
\begin{equation}
V(k,\beta) = \displaystyle\sum_{\alpha,\;\gamma} \; v^{\alpha,\gamma}_{k,\beta} \; \psi^\dagger(k; \alpha) \; \psi(k;\gamma) \;.
\label{eq-1}
\end{equation} 
Here, $\alpha$ and $\gamma$ are $k$-particle configuration states in occupation number basis. Distributing $k$ particles (fermions in agreement with Pauli's exclusion principle or bosons) in $N$ sp states will generate the complete set of these distinct configurations. Total number of these configurations are $\binom{N}{k}$ for fermions and $\binom{N+k-1}{k}$ for bosons. In occupation number basis, we order the sp levels (denoted by $\mu_i$) in increasing order, $\mu_1 \leq \mu_2 \leq \cdots \leq \mu_N$. Operators $\psi^\dagger(k; \alpha)$ and $\psi(k;\gamma)$ respectively are $k$-particle creation and annihilation operators for fermions or bosons, i.e. $\psi^\dagger(k; \alpha) = \prod_{i=1}^{k} a^\dagger_{\mu_i}$, $\psi(k;\gamma) = \prod_{i=1}^{k} a_{\mu_i}$ for fermions and $\psi^\dagger(k; \alpha) =  {\cal N}_{\alpha} \; \prod_{i=1}^{k} b^\dagger_{\mu_i}$, $\psi(k;\gamma) = {\cal N}_{\gamma} \; \prod_{i=1}^{k} b_{\mu_i}$ for bosons. Here, ${\cal N}_{\alpha}$ and ${\cal N}_{\gamma}$ are the factors that guarantees unit normalization of $k$-particle bosonic states. The sum in Eq. \eqref{eq-1} stands for summing over a subset of $k$-particle creation and annihilation operators.
These $k$-particle operators obey the usual anti-commutation (commutation) relations for fermions (bosons). 

In Equation \eqref{eq-1}, $v^{\alpha,\;\gamma}_{k,\beta}$ is chosen to be a $\binom{N}{k}$ [$\binom{N+k-1}{k}$] dimensional GOE/GUE (depending on $\beta$ value) in $k$-particle spaces. That means $v^{\alpha,\;\gamma}_{k,\beta}$ are anti-symmetrized (symmetrized) few-body matrix elements for fermions (bosons) chosen to be randomly distributed independent Gaussian variables with zero mean and variance
\begin{equation}
{\overline{v^{\alpha,\gamma}_{k,\beta} \; v^{\alpha^\prime,\gamma^\prime}_{k,\beta}}} = v^2 \; \left( {\delta_{\alpha,\gamma^\prime}} {\delta_{\alpha^\prime,\gamma}} + \delta_{\beta,1} {\delta_{\alpha,\alpha^\prime}} {\delta_{\gamma^\prime,\gamma}} \right) \;.
\label{eq-2}
\end{equation} 
Here, the bar denotes ensemble averaging and we choose $v=1$ without loss of generality. 

Distributing the $m$ fermions (bosons) in all possible ways in $N$ levels generates the many-particle basis states defining $d_F(N,m)=\binom{N}{m}$ [$d_B(N,m)=\binom{N+m-1}{m}$] dimensional Hilbert space. The action of the Hamiltonian operator $V(k,\beta)$ defined by Equation \eqref{eq-1} on the many-particle states generates the FEGOE($k$)/FEGUE$(k)$/BEGOE($k$)/BEGUE$(k)$ ensemble in $m$-particle spaces.

\section{Formulas of $q$}
\label{sec2}

\subsection{FEGUE($k)$ and FEGOE($k$)}

In the dilute limit defined by $N \rightarrow \infty$, $m \rightarrow
\infty$, $k \rightarrow \infty$, $m/N \rightarrow 0$ and $k/m \rightarrow 0$,
the moments $\lan H^p \ran^m$ of the spectral density generated by FEGOE($k$)
and FEGUE($k$) will be those of a Gaussian. This result is easily derived and
well known \cite{MF,kota,BW,Mu-16}. The reduced moments to order eight with $N
\rightarrow \infty$, $m \rightarrow \infty$ but $k/m$ finite are derived in 
\cite{MF} for FEGOE($k$) and in \cite{Mu-16} for FEGUE($k$). For example, 
for FEGUE($k$), the reduced moments up to order 8 are 
\cite{Mu-16},
\be
\barr{l}
\mu_4(m,k) =  2+G(m,k,1) \;,\\ \\
\mu_6(m,k) =  5 + 6 G(m,k,1)+3 \l[G(m,k,1)\r]^2 
\\
+ G(m,k,2)G(m,k,1)\;,\\ \\
\mu_8(m,k) = 14+28G(m,k,1) + 28\l[G(m,k,1)\r]^2 
\\ 
 +  12\l[G(m,k,1)\r]^3 + 8G(m,k,2)G(m,k,1) \\
 +  4G(m,k,1) \l[G(m,k,2)\r]^2 \\
 +  8 \l[G(m,k,1)\r]^2 G(m,k,2)\\
 +  G(m,k,1)G(m,k,2)G(m,k,3) \\
 +  2 \l[G(m,k,1)\r]^2 \dis\sum_\alpha 
\dis\frac{\dis\binom{k}{\alpha}^2 \dis\binom{m-2k}{k-\alpha}}{\dis\binom{m}{k} 
\dis\binom{m-k}{\alpha}}\;; \\ \\
G(m,k,r)=\dis\frac{\dis\binom{m-rk}{k}}{\dis\binom{m}{k}}\;.
\earr \label{eq.qh7}
\ee
Comparing these with the FEGOE($k$) formulas given in \cite{MF}, it is seen that
the moments to order 6 for FEGOE($k$) are same as those given in Eq.
(\ref{eq.qh7}) and for $\mu_8$ only the last term is different.  Comparing Eqs.
(\ref{eq.qh7}) and (\ref{eq.qh6}), it is seen that the lower order reduced
moments of FEGUE($k)$ and FEGOE($k$) will be essentially same as those of the
generating function of $q$-Hermite polynomials with $q$ given by
\be
q \sim G(m,k,1) = \dis\frac{\dis\binom{m-k}{k}}{\dis\binom{m}{k}} = \mu_4-2\;.
\label{eq.qh8}
\ee
Using finite-$N$ corrections to $\mu_4$ as given in \cite{BW,Ko-05,kota}, a 
better approximation for $q$ for FEGUE($k$) is
\be
\barr{l}
q \sim \dis\binom{N}{m}^{-1} \dis\sum_{\nu=0}^{\nu_{max}}\; 
\dis\frac{\Lambda^\nu(N,m,m-k)\;\Lambda^\nu(N,m,k)\;d(g_\nu)}{
\l[\Lambda^0(N,m,k)\r]^2} \,; \\ \\
\Lambda^\nu(N,m,r) =  \dis\binom{m-\nu}{r}\;\dis\binom{N-m+r-\nu}{r}\;,\\ \\
d(g_\nu)  = \dis\binom{N}{\nu}^2-\dis\binom{N}{\nu-1}^2\;.
\earr \label{eq.qh9}
\ee
Here $\nu_{max} = $ min$\{k,m-k\}$ in the summation.
On the other hand, using finite $N$ corrections to $\mu_4$ as given in 
\cite{Man-th,kota} for FEGOE($k$), a better approximation for $q$ for 
FEGOE($k$) is,
\be
\barr{l}
q \sim F(N,m,k)/\l[T(N,m,k)\r]^2 \\ \\
T(N,m,k)= \dis\binom{m}{k}\l[\dis\binom{N-m+k}{k} + 1 \r]\;,\\ \\
F(N,m,k)= \dis\binom{m}{k}^2 +  \dis\sum_{s=0}^{k} \dis\binom{m-s}{k-s}^2
\dis\binom{N-m+k-s}{k} \\ \\
\times \dis\binom{m-s}{k} \dis\binom{N-m}{s}
\dis\binom{m}{s} \\ \\
\times \l[\dis\frac{N-2s+1}{N-s+1}\r]
\dis\binom{N-s}{k}^{-1} \dis\binom{k}{s}^{-1}\,\l\{2+\dis\binom{N+1}{s}\r\} \;.
\earr \label{eq.qh9a}
\ee
Eqs. (\ref{eq.qh8}), (\ref{eq.qh9}) and (\ref{eq.qh9a}) show that the expression 
for $q$ mentioned in \cite{Verb-3} is inappropriate as it has no particle number ($m$) dependence. Table \ref{tab.1} gives some numerical results for $q$ as $k$ changes. Note that the numerical values of $q$ in these examples are same for FEGOE($k$) and FEGUE($k$) up to three decimal places. As we see from the table, the $q$ becomes zero rapidly with increasing $k$ even for a very large system.

\begin{table}
\caption{Values of parameter $q$ for FEGOE($k$)[FEGUE($k$)] for various $(N,m)$ as a function of rank of interaction $k$ computed using Eq. \eqref{eq.qh9a}[\eqref{eq.qh9}].}
\label{tab.1}
\begin{center}
\scalebox{1.0}{
\begin{tabular}{cccccccc}
\hline
$N$  & $m$ & $k$ & $q$ & $N$ & $m$ & $k$ & $q$ \\ \hline
 12 & 6  & 1  & 0.735 & & & 4 & 0.015 \\
 & & 2 & 0.287 &  & & 5 & 0.001 \\
 & & 3 & 0.057 &  & & $\geq$ 6 & 0.000 \\
 & & 4 & 0.005 & 50 & 10 & 1 & 0.879 \\
 & & $\geq$ 5 & 0.000 & & & 2 & 0.567 \\
20 & 8 & 1 & 0.814 &  & & 3 & 0.239 \\
 & & 2 & 0.417 &  & & 4 & 0.053 \\
 & & 3 & 0.119 &  & & 5 & 0.003  \\
 & & & & & & $\geq$ 6 & 0.000 \\ \hline
\end{tabular}}
\end{center}
\end{table}

\subsection{BEGUE($k$) and BEGOE$(k)$}

\begin{figure}
\begin{tabular}{cc}
\includegraphics[width=3.2in,height=6in]{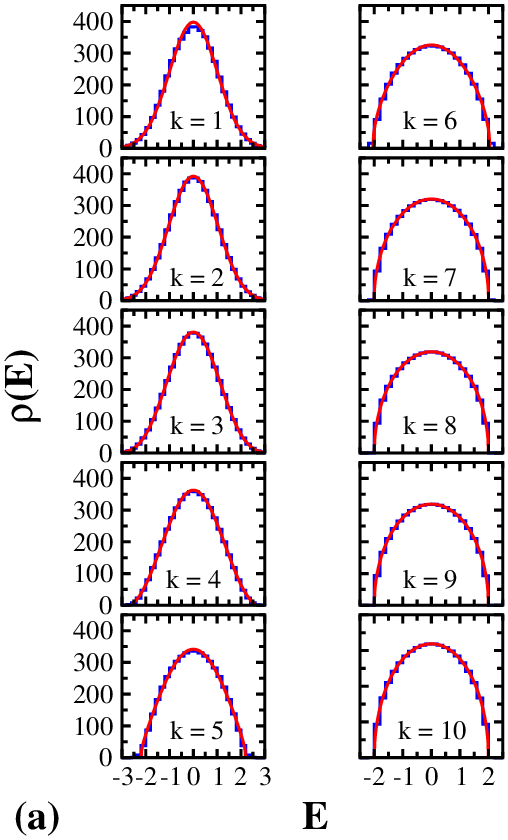} &
\includegraphics[width=3.2in,height=6in]{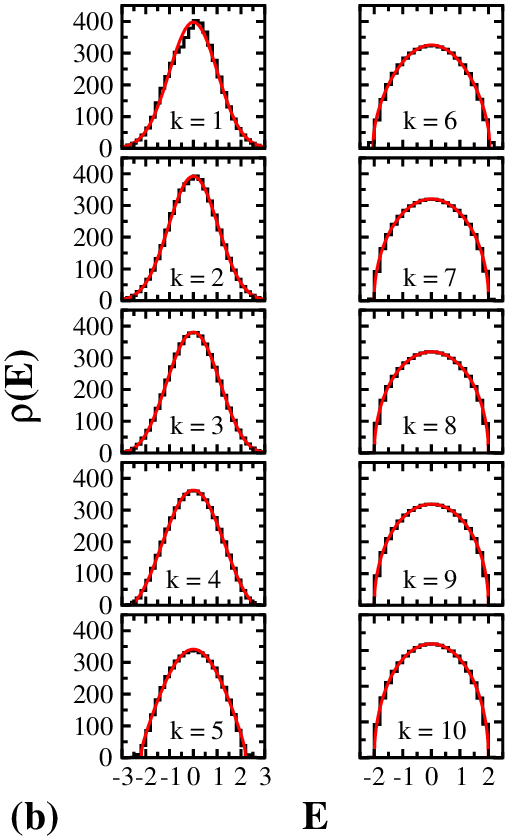} \\
\end{tabular}
\caption{(color online) Ensemble averaged spectral density (histograms) of a 1000 member (a) BEGUE(k) and (b) BEGOE$(k)$ with $N=5$ and $m=10$ and $k$ changing from 1 to 10. The smooth (red) curves are obtained using Eq. \eqref{eq.qh9b} with the corresponding values of $q$ given in Table \ref{tab.2}.}
\label{den-begoek}	
\end{figure}

\begin{figure}
\includegraphics[width=3.2in,height=5in]{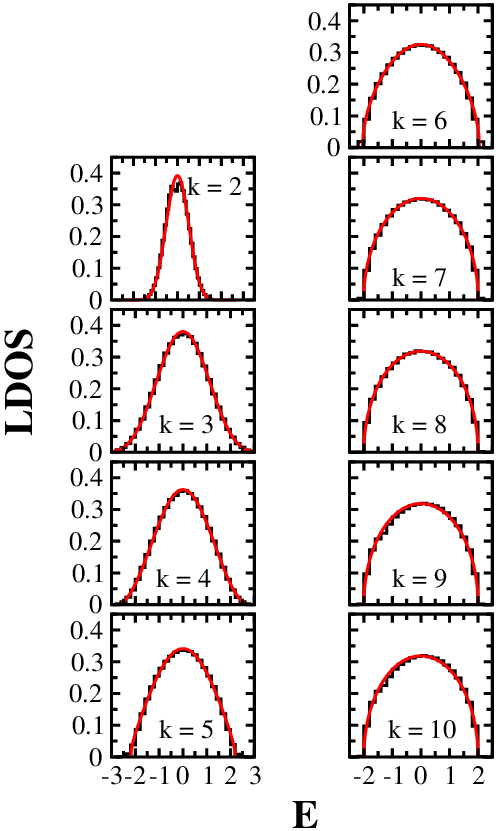}
\caption{(color online) Ensemble averaged LDOS of a 1000 member BEGOE$(1+k)$ (histograms) with $N=5$ and $m=10$ and $k$ changing from 2 to 10. The smooth (red) curves are obtained using Eq. \eqref{eq.qh4} with the corresponding values of $q$ given in Table \ref{tab.2}.}
\label{sfunc-begoek}	
\end{figure}

Formula for $\mu_4$ for BEGUE($k$), with finite $N$ corrections, follows easily
from Eq. (\ref{eq.qh9}) by using the so called $N \rightarrow -N$ law as 
described in \cite{Ko-05} and confirmed by the explicit derivation given in
\cite{Asaga}. Then the formula for $q$ for BEGUE($k$) is,
\be
\barr{l}
q \sim \dis\binom{N+m-1}{m}^{-1} \dis\sum_{\nu=0}^{\nu_{max}}\; \dis\frac{
X(N,m,k,\nu) \;d_B(g_\nu)}{\l[\Lambda_B^0(N,m,k)\r]^2}
\\ \\
X(N,m,k,\nu) = \Lambda_B^\nu(N,m,m-k)\;\Lambda_B^\nu(N,m,k) ; \\ \\
\Lambda_B^\nu(N,m,r) =  \dis\binom{m-\nu}{r}\;\dis\binom{N+m+\nu-1}{r}\;,\\ \\
d_B(g_\nu)  = \dis\binom{N+\nu-1}{\nu}^2-\dis\binom{N+\nu-2}{\nu-1}^2\;.
\earr \label{eq.qh9b}
\ee
Some numerical results for $q$ calculated using Eq. (\ref{eq.qh9b}) are given in
Table \ref{tab.2}. Unlike the situation with FEGOE($k$), formula for $\mu_4$ for
BEGOE($k$), for a general $k$ value, is not available \cite{Asaga}. However, it
is expected that Eq. (\ref{eq.qh9b}) gives a good approximation to $q$ for
BEGOE($k$) and this is confirmed with numerical calculations in Figs. \ref{den-begoek} and \ref{sfunc-begoek} ahead.

\begin{table}
\caption{Values of parameter $q$ for BEGUE($k$)/BEGOE($k$) for various $(N,m)$ as a function of rank of interaction $k$ computed using Eq. \eqref{eq.qh9b}.}
\label{tab.2}
\begin{center}
\begin{tabular}{cccccccc}
\hline
$N$  & $m$ & $k$ & $q$ & $N$  & $m$ & $k$ & $q$ \\ \hline
 5 & 10 & 1 & 0.969 &  & & 4 & 0.712 \\
 & & 2 & 0.861 &  & & 5 & 0.556 \\
 & & 3 & 0.664 &  & & 6 & 0.392 \\
 & & 4 & 0.405 &  & & 7 & 0.242 \\
 & & 5 & 0.172 &  & & 8 & 0.127 \\
 & & 6 & 0.045 &  & & 9 & 0.054 \\
 & & 7 & 0.007 &  & & 10 & 0.018 \\
 & & $\geq$ 8 & 0.000 &  & & 11 & 0.005 \\
 10 & 20 & 1 & 0.984 &  & & 12 & 0.001 \\
 & & 2 & 0.932 &  & & $\geq$ 13 & 0.000 \\
 & & 3 & 0.840 & & & & \\ \hline
\end{tabular}
\end{center}
\end{table}

\begin{figure}
\begin{tabular}{cc}
\includegraphics[width=3.25in,height=4in]{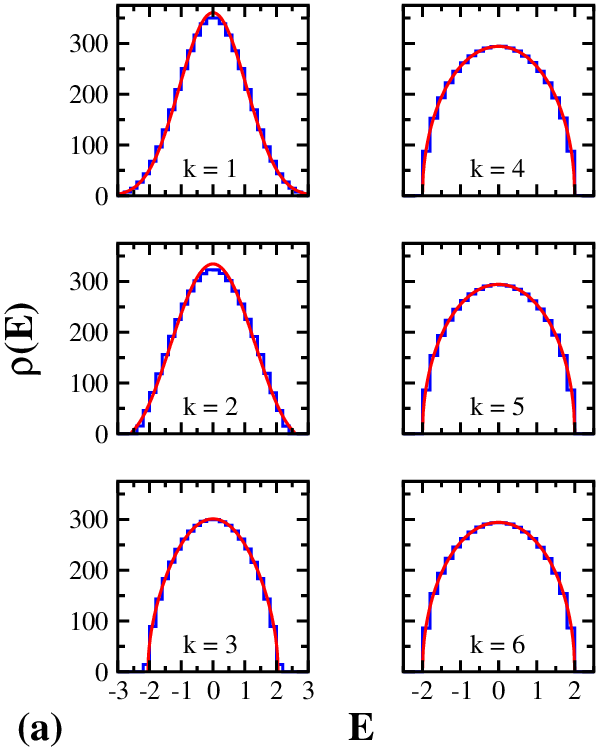} &
\includegraphics[width=3.25in,height=4in]{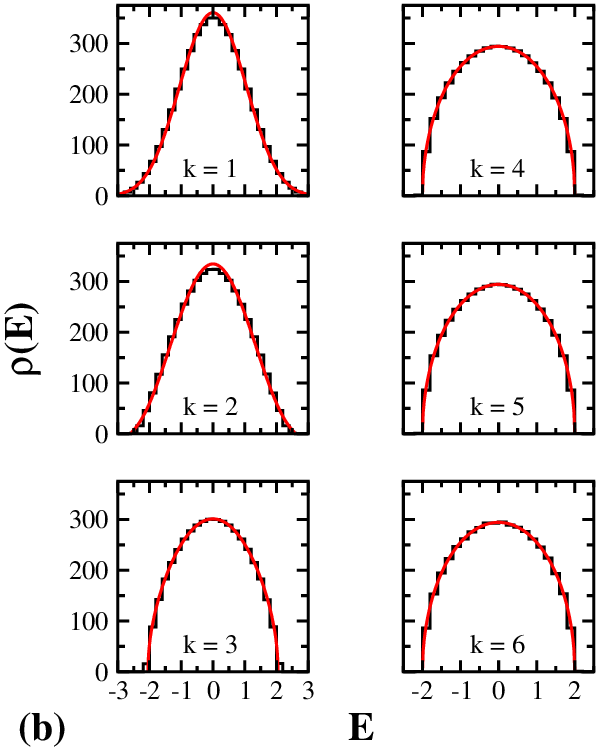} \\
\end{tabular}
\caption{(color online) Ensemble averaged spectral density (histograms) of a 1000 member (a) FEGUE($k$) and (b) FEGOE$(k)$ with $N=12$ and $m=6$ and $k$ changing from 1 to 6. The smooth (red) curves are obtained using Eq. \eqref{eq.qh10} with the corresponding values of $q$ given in Table \ref{tab.1}.}
\label{den-fegoek}	
\end{figure}

\section{Gaussian to semi-circle transition in spectral density}
\label{sec3}

Normalizing the eigenvalues $E$ with centroid $E_c$ and spectrum width
$\sigma$, we have ${\bf E}=(E-E_c)/\sigma$. Then the spectral density $\rho(E)dE$
for the four EE($k$) ensembles considered, from the above results, is given by
\be
\barr{l}
\rho(E) dE = dE\;\can_q\,\dis\frac{1}{\sigma}\dis\sqrt{1-\dis\frac{(E-E_c)^2}{
E_0^2}}\\ \\
\times 
\dis\prod^{\infty}_{\kappa=1} \l[
1-\dis\frac{4\l\{(E-E_c)^2/E_0^2\r\}}{2+q^\kappa+q^{-\kappa}}\r]\;;\\ \\
E_0^2=\dis\frac{4\sigma^2}{1-q}\;,\;\;
\dis\frac{E_c-2\sigma}{\sqrt{1-q}} \leq E \leq \dis\frac{E_c+2\sigma}{
\sqrt{1-q}}\;.
\earr \label{eq.qh10}
\ee
Limits in this equation show that the density will go to zero at $E_c \pm
2\sigma/\sqrt{1-q}$. This gives correctly the known results for Gaussian ($q=1$)
and semi-circle ($q=0$). The infinite product in Eq. (\ref{eq.qh10}) can be 
simplified in some situations \cite{Verb-2,bh-1}. Also, it is easy to write the
formulas for the spectral variance $\sigma^2$ for the four ensembles
\cite{kota}. Eq. (\ref{eq.qh10}) along with the formulas  for $q$ is tested in
some numerical calculations and the results are shown in Figures \ref{den-fegoek} and \ref{den-begoek}. 

Figure \ref{den-fegoek} shows the ensemble averaged spectral density of a 1000 member FEGOE($k$) and FEGUE($k)$ (histograms) with $N=12$ and $m=6$ as a function of $k$. Corresponding Hamiltonians are $924$ dimensional. The smooth curves are obtained using Eqs. \eqref{eq.qh10}, \eqref{eq.qh9} and \eqref{eq.qh9a} with the  corresponding values of $q$ given in Table \ref{tab.1}. As seen from the figure, the results for FEGOE$(k)$ and FEGUE$(k)$ are identical for all $k$ values. Eigenvalue density is Gaussian for $k=1$ and 2 and makes a transition to semi-circle at $k=3$. The agreement between the theory given by the $q$-Hermite polynomials and numerics is excellent.

Similarly, Figure \ref{den-begoek} shows transition in spectral density of a 1000 member BEGOE$(k)$ and BEGUE($k)$ (histograms) with $N=5$ and $m=10$ as a function of $k$. Corresponding Hamiltonians are $1001$ dimensional. The smooth curves are obtained using Eqs. \eqref{eq.qh9b} and \eqref{eq.qh10} with the corresponding values of $q$ given in Table \ref{tab.2}. Here, the results for BEGOE($k$) and BEGUE($k$) are same except for some deviations for $k = 1$. Eigenvalue density is Gaussian till $k=4$ and makes a transition to semi-circle at $k = 5$. The agreement between the theory given by $q$-Hermite polynomials and numerics is excellent. 

\begin{figure}
\includegraphics[width=3.2in,height=5in]{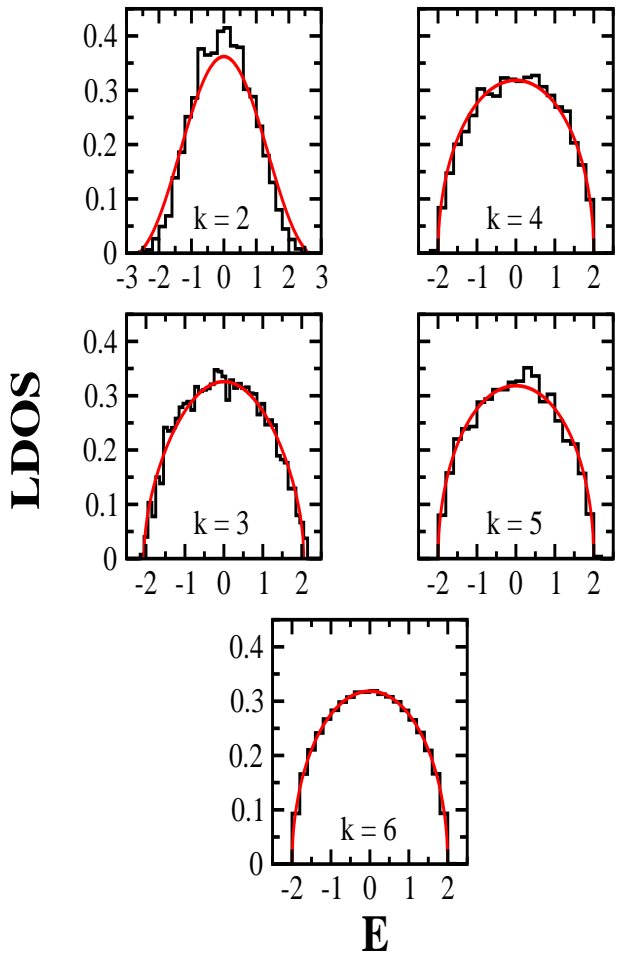}
\caption{(color online) Ensemble averaged LDOS of a 1000 member FEGOE$(1+k)$ (histograms) with $N=12$ and $m=6$ and $k$ changing from 2 to 6. The smooth (red) curves are obtained using Eq. \eqref{eq.qh4} with the corresponding values of $q$ given in Table \ref{tab.1}.}
\label{sfunc-fegoek}	
\end{figure}

\begin{figure}[ht]
\includegraphics[width=4.5in]{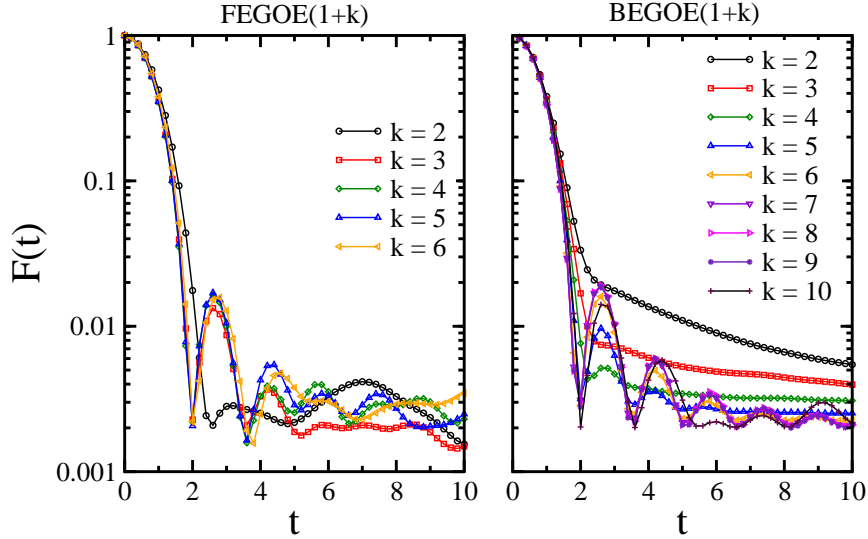}
\caption{(color online) Survival probability decay $F(t)$ for: (left panel) a 1000 member FEGOE$(1+k)$ ensemble with $N=12, \; m=6$ and (right panel) a 1000 member BEGOE$(1+k)$ ensemble with $N=5, \; m=10$. Results are shown for various values of $k$.}
\label{sprob-comp}	
\end{figure}

\section{LDOS and survival probability decay in FEGOE$(1+k)$ and BEGOE$(1+k)$}
\label{sec4}

Understanding non-equilibrium dynamics of interacting many-body quantum systems is fundamental for many branches of physics \cite{zel-lea, Mal-16, Mu-09}. Unitary evolution of quantum systems is investigated experimentally using cold atoms, ion traps and nuclear magnetic resonance \cite{Ka-16, Sm-16, Ga-16, Wei-16}. In order to characterize system evolution, we analyze relaxation dynamics of survival probabilities. Survival probability $F(t)$ is the probability to find the system still in the initial state $\l|\psi(0)\ran$ after time $t$,
\be
F(t) = \l|\lan \psi(0) | \exp(-i H t) | \psi(0) \ran \r|^2 \;.
\label{eq-sp}
\ee
The system is prepared in an eigenstate $\l|\psi(0)\ran$ of unperturbed mean-field Hamiltonian $h(1)$. Dynamics starts with a sudden change in the parameter $\lambda$ (denoting strength of perturbation) in a time interval much shorter than any of the characteristic time scale of the model. With a quench $V(k)$ (we consider only $\beta=1$ and hence, drop it from now on) of strength $\lambda$, this results in a final (perturbed) Hamiltonian 
\be
H = h(1) + \lambda \; V(k) 
\label{eq-hpv}
\ee
with eigenvalues $E$ and eigenstates $\l|E\ran \neq \l|0\ran$. The initial state 
$\l|\psi(0)\ran$ unitarily changes after time $t$ as $\psi(t) = \exp(-i H t) \l|\psi(0)\ran$.

Expanding the mean-field initial state $\l|\psi(0)\ran$ over the eigenstates $\l|E\ran$, LDOS is defined as,
\be
\mbox{LDOS} = \dis\sum_{E^\pr} \l| C_{\psi(0)}^{E^\pr} \r|^2 \delta(E-E^\pr) \;,
\label{eq-ldos}
\ee
with $C_{\psi(0)}^E = \lan E | \psi(0) \ran$ being the overlaps. Thus, LDOS give the spread of basis states over the eigenstates. 

Equation \ref{eq-hpv} denotes FEGOE(1+$k$) or BEGOE(1+$k$) depending on the choice of particles (fermions or bosons) with $V(k)$ defined by Eq. \ref{eq-1}. The mean-field Hamiltonian $h(1)$ is defined by the fixed sp energies $\epsilon_i = i +1/i$. We choose $\lambda=0.5$ close to the region of thermalization \cite{Ma-11,kota}. We construct 1000 member FEGOE$(1+k)$ ensemble with $N=12$, $m=6$, $k$ varying from $2-6$ and 1000 member BEGOE$(1+k)$ ensemble with $N=5$, $m=10$, $k$ varying from $2-10$. Corresponding matrix dimensions are $d_F(12,6)=924$ and $d_B(5,10)=1001$. LDOS are then computed as follows. First of all, the basis state energies $e_b(m)$ are the diagonal elements of the $H$ matrix in the many-particle Fock-space basis giving $e_b(m) = \lan b | h(1) + \lambda V(k) | b \ran$. Note that the centroids of the $e_b(m)$ energies are the same as that of the eigenvalue $(E)$ spectra but their widths are different. For each member of the ensemble, energies $E$ and $e_b(m)$ are zero centered and scaled by the spectrum width $\sigma_H(m)$. For each member of the ensemble, $\l|C_{\psi(0)}^E\r|^2$ are summed over the basis states in the energy windows $e_{\psi(0)}(m) \pm \delta$ and then ensemble averaged LDOS are constructed as histograms as a function of energy. We choose $e_{\psi(0)}(m)=0$ and $\delta=0.2$. Results are shown in Figs. \ref{sfunc-fegoek} and \ref{sfunc-begoek}. For sufficiently strong $k$-body interactions, the LDOS change from Gaussian to semi-circle form just as the spectral density. Importantly,  LDOS follow $v(x|q)$ as $k$ changes from $2$ to $m$. Agreement between the numerical histograms and solid curves obtained using Eq. \ref{eq.qh4} is excellent.

Using Eq. \ref{eq-sp}, we easily see that survival probability is essentially given by the Fourier transform of the LDOS,
\be
\barr{rcl}
F(t) & = & \l| \dis\sum_E  \l| C_{\psi(0)}^E \r|^2 \exp{-iEt} \r|^2 \\ \\
 & = &  \l| \dis\int  \mbox{LDOS} \; \exp{-iEt} \; dE \r|^2 \;.
\earr
\label{eq.fid}
\ee 
Survival probabilities are computed as follows. For each member at a given time $t$, $|\sum_E  C_{\psi(0)}^E C_{\psi(0)}^E|^2$ are summed over the basis states $\l|\psi(0)\ran$ in the energy window $e_{\psi(0)}(m) \pm \delta_1$. We choose $\delta_1= 0.01$ and $e_{\psi(0)}(m) = 0$. Then, ensemble averaged survival probability for a fixed initial mean-field basis state is obtained by binning. 

\begin{figure}
\includegraphics[width=3.15in,height=5.25in]{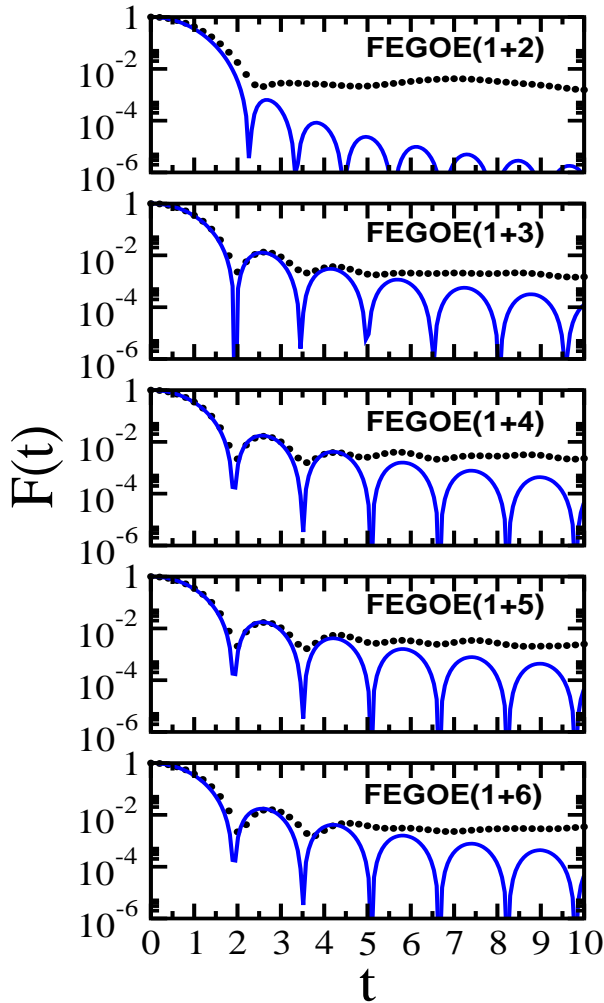}
\caption{(color online) Survival probability decay $F(t)$ for a 1000 member FEGOE$(1+k)$ ensemble with $N=12$ and $m=6$ as a function of $k$. Monte-Carlo results for the survival probability decay (filled circles) are compared with the  numerical Fourier transform of Eq. \eqref{eq.qh10} (solid (blue) curves).}
\label{sprob-fegoek}	
\end{figure}

\begin{figure}
\includegraphics[width=3.15in,height=5.25in]{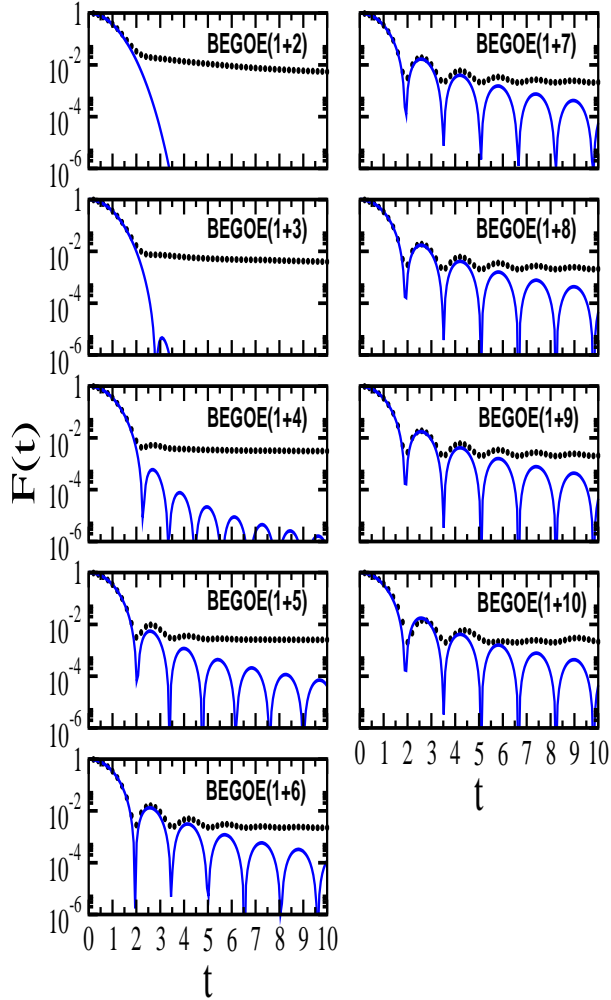}
\caption{(color online) Survival probability decay $F(t)$ for a 1000 member BEGOE$(1+k)$ ensemble with $N=5$ and $m=10$ as a function of $k$. Monte-Carlo results for the survival probability decay (filled circles) are compared with the numerical Fourier transform of Eq. \eqref{eq.qh10} (solid (blue) curves).}
\label{sprob-begoek}	
\end{figure}

Monte-Carlo results for survival probability decays for FEGOE(1+$k$) and BEGOE(1+$k$) as a function of $k$ are shown in Fig. \ref{sprob-comp}. Decay becomes faster with increasing $k$ for both FEGOE(1+$k$) and BEGOE(1+$k$). Survival probability decay shows oscillations for all $k$ values in FEGOE(1+$k$) while for BEGOE(1+$k$), oscillations are seen only for $k \geq 4$ and they become more pronounced with increasing $k$. Analytical result for the Fourier transform of $v(x|q)$ are not yet available (here, results in \cite{Fourier} may prove to be useful). Therefore, we have numerically computed the Fourier transform of Eq. \eqref{eq.qh10} and compared it  with the Monte-Carlo results for the survival probabilities in Figs. \ref{sprob-fegoek} and \ref{sprob-begoek}.

For a 1000 member FEGOE$(1+k)$ ensemble with $N=12$ and $m=6$, we compare the Monte-Carlo results for survival probability decay (solid circles) with numerical Fourier transform of Eq. \eqref{eq.qh10} in Figure \ref{sprob-fegoek}. Crossover to the region of thermalization will be faster with increasing $k$; see discussion in Appendix G of \cite{kota} for a comparison between FEGOE(1+2) and FEGOE(1+3). Due to this, the results for survival probability for $k=4-6$ are same. As seen from the figure, results for survival probability show oscillations with increasing time $t$. The Fourier transform of Eq. \eqref{eq.qh10} describes the short-time behavior accurately and the agreement gets better with increasing $k$. It also captures the positions of the oscillations. It is important to note that the oscillations are damped for Monte-Carlo calculations because of two reasons: (i) we approximate the LDOS appearing in Eq. \eqref{eq.fid} by $v(E|q)$ which represents only the smoothed part of LDOS; and (ii) there is averaging of LDOS in a window of width $\delta=0.2$ around $e_{\psi_0}(m)=0$. Further investigations of this will be reported elsewhere.

Similarly, Figure \ref{sprob-begoek} shows comparison of Monte-Carlo results for survival probability decay (solid circles) with numerical Fourier transform of Eq. \eqref{eq.qh10} for a 1000 member BEGOE$(1+k)$ ensemble with $N=5$, $m=10$ and $\lambda = 0.5$. As the crossover to region of thermalization is faster in bosonic systems compared to fermionic ones \cite{kota}, results for $k = 7-10$ are same. Decay shows oscillations with increasing $t$ and these become prominent with increasing $k$. Short time dynamics and positions of oscillations are well captured by Fourier transform of Eq. \eqref{eq.qh10}. 

\section{Conclusions}
\label{sec5}

We have shown that the generating function of $q$-Hermite polynomials describes the Gaussian to semi-circle transition in the spectral densities and LDOS of FEGOE$(k)$ / FEGUE$(k)$ / BEGOE$(k)$ / BEGUE($k$) as a function of rank of interactions $k$ by deriving the formula for $q$ parameter for these few-body ensembles. Survival probability decay in FEGOE$(1+k)$ and BEGOE$(1+k)$ is explained by the numerical Fourier transform of the generating function of $q$-Hermite polynomials. Dynamics of non-equilibrium quantum systems depends strongly on the nature of particles (fermions or bosons) and the rank of interactions $k$. Although two body interactions are dominant, three and higher body interactions may become prominent in strongly interacting quantum systems \cite{Bl-75, RMP-12}. The present results establish that $q$-Hermite polynomials (also may be other $q$-special functions) play an important role in embedded ensembles in explaining the dependence of spectral density, LDOS and survival probabilities on $k$ and the nature of particles. In future, it will be good to explore bivariate $q$-Hermite polynomials and their generating function as they may prove to be useful understanding the two-point correlation functions that determine level fluctuations and also transition strengths generated by a transition operator \cite{kota, Sz-10}.

\section*{Acknowledgments} 

M. V. acknowledges financial support from UNAM/DGAPA/PAPIIT research grant IA101719. We thank N. D. Chavda for discussions.

\ed